\documentclass[12pt]{article}
\pdfoutput=1
\usepackage[utf8]{inputenc}     
\usepackage{microtype}           
\usepackage{graphicx}             
\usepackage{xcolor}             
\usepackage{booktabs}			 
\usepackage{tabularx}              
\usepackage{adjustbox}           
\usepackage{hyperref}			 

\usepackage{lipsum}
\usepackage{enumitem}
\setlist[enumerate]{itemsep=0mm}

\AtBeginDocument{
    \label{CorrectFirstPageLabel}
    
}

\usepackage{authblk}
\author[1]{Sandro Ferreira Sousa}
\author[1]{Camilo Rodrigues Neto\footnote{corresponding author: camiloneto@usp.br}}
\author[1]{Fernando Fagundes Ferreira}

\affil[1]{EACH, São Paulo University (USP), Av. Arlindo Bettio 1000, 03828-000, São Paulo, SP, Brasil}

\begin{document}

\title{Structure and robustness of São Paulo public transport network}

\maketitle

\begin{abstract}
{Public Transport Networks plays a central role in cities devolopment. In big cities such system may be represented by a complex network and understand its properties is of great interest for managers and scholars.
In this work, the urban public transport system of São Paulo is reinterpreted as a coupled (bus, subway and train) network, bypassing operational details and focusing on connectivity. 
Using a empirically generated graph, a statistical characterization is made by network metrics.
Nearby bus stops and rail transport stations (subway and train) may or may not be considered as a single vertex in the network representation of the transport system, depending on how much an user is willing to walk to shift from one stop/station to another.
This distance radius is then used to group nearby stops/stations as a single vertex in the network representation of the urban public transport system and then its properties are studied as a function of this radius.
This radius is used as proxy of the user's willingness to walk until the nearest point to access transportation. 
The variation of the measure $ \rho $ leads to changes in the perception of the topology of the public transport network as shown in this work. An interesting result was that the network is assortative. Another aspect investigated was the degree distribution of the network. It was not possible to distinguish between power-law or a log-normal distribution.
An exploratory model is used to test the robustness of the network by randomly, deterministically and preferentially targeting the stops and service lines. 
According to the grouping radius, aka willingness, different fragmentation values were obtained under attack simulations.
We showed that increasing this willingness generates great reduction in the number of necessary jumps between buses, subway and trains lines to achieve all the network destinations.}
{Complex Networks, Public Transportation, Robustness}
\end{abstract}

\section{Introduction}

Public Transport Networks (PTN) are constantly experiencing interruptions due to several causes, as unintended mechanical failures or intended attacks. These events can result on large-scale impact to the city causing delays, financial losses  and  jeopardizing  passenger routine. The stability of this type of critical infrastructure is vital as it plays an important role in the city dynamics and mobility flows \cite{batty2013new}. Finding the most important (or vulnerable) elements in PTNs and making them more robust is a useful way to establish more efficient and reliable systems. In this context, robustness is understood as the ability of the network to absorb disturbances to nodes (stops) or links (service lines) and to continue operating in nearly the same conditions found in a normal situation, that is, connecting users to their destinations \cite{rodriguez2014measuring}.

In practice, dealing with PTNs in big cities is not trivial. They show a complex combination of buses, subway and urban trains commonly managed by different agencies. 
This is the case of Sao Paulo, the biggest Brazilian city in the fifth largest urban agglomeration in the world with more than $21$ millions inhabitants in the metropolitan area \cite{united2014world}. 
The bus system consist of more than $1,300$ service lines covering the city area while subway and train combined spread around the metropolitan area with $11$ lines and $153$ stations. The impact of a stop in the transportation system can be huge, as nearly half the area’s population uses it daily to commute with more than $9.5$ million passengers transported per working day by this network.

There is no doubt that public transportation plays a crucial role in urban regions of our modern society. In recent years, several transport systems have been investigated with the framework of network science. 
A group of researchers compared the PTNs of Berlin, Düsseldorf and Paris finding the scale-free property \cite{barabasi2000scale} in these three networks \cite{von2005scaling}. A sequent work compared statistical properties of $14$ PTNs of large cities around the world, considering their geographic information in different spatial representations \cite{von2007network}. A growth model based on simple rules was proposed to support the statistical properties found in the same set of networks \cite{von2009public}.

Similarly, the small-world behavior \cite{watts1998collective} was found for 22 PTNs in Poland and the degree distribution of stops follows a power law and exponential function according to the approach for space representation \cite{sienkiewicz2005statistical}. The same phenomena was found for the PTN in Boston, where an efficiency measure was proposed to deal with transport networks \cite{latora2002boston}. Beijing and Chengdu in China produced a similar power law for the degree distribution while a weight based on passengers flow was used for links between stations \cite{Ma2011power}. This phenomena was confirmed on a larger study in China with 330 cities using the $P$ space (Transfer Space), but in contrast, the degree distribution of stops was described by an exponential function, sugesting a random growth network mechanism \cite{qing2013space}.

These statistical properties showed an interesting finding about PTNs: they appear robust under random failure but vulnerable when the node plays an important role in the network, such as a hub \cite{von2007attack}. A study on 33 different subway systems around the world performed a robustness analysis using an approach similar to $P$ space \cite{derrible2010complexity}. They found the characteristic small-world behavior and a higher robustness in networks more clustered, that is, redundancies increase robustness. For the PTNs in London and Paris, a quantitative analysis was performed comparing which topology presented higher robustness in the $L$ space (Geographical Stop Space), observing how removing lines or stops affect the overall functionality \cite{von2012tale}. The degree distribution could be described by a power law in both networks, but Paris proved to be more robust than London. In addition, the study proposed the investigation of clusters formation and the addition of $P$ space representation, permitting the observation of cascade effect on routes by removing stops.

The small-world phenomenon frequently discussed in transport networks is interesting because it presents a topology with proportions that refer to robustness and efficiency for the system. That is, a high clustering (redundancies) and a short average path length to connect them, in which, commonly exhibit short-cuts to reach great distances in few steps. The scale-free property is characterized by the presence of highly connected nodes (hubs), consisting of tens, hundreds or thousands of links. In this sense, the network seems to have no scale \cite{barabasi2016network}. However, removing these hubs can cause greater impact on the network as a whole, making it vulnerable.

The goal of this paper is to study São Paulo's PTN as a coupled system (bus and rails) by a descriptive analysis of its statistical properties in network metrics \cite{newman2003structure, boccaletti2006complex}. 
In section~\ref{empirical}, empirical data is used to create the network where different distance radius values are used to group nearby stops and stations. This grouping could be useful as a measure for public policy and may represent the user's willingness to move to nearby  stops for transportation access. 
Further, in section~\ref{attacks}, a data-based exploratory model is formulated to tests the network robustness. Attack simulations are performed aiming stations and lines to observe the system ability to keep connecting destinations in the face of ruptures.

\section{Empirical analysis}\label{empirical}

The PTN of São Paulo is investigated as a coupled network, that is, bus, subway and train lines are interconnected, but without any distinction between them. The stations, terminals or stops are represented as nodes and their lines and routes as links according to its direction. This approach is well known as $L$ space \cite{goh2014emergence, von2012tale, latora2002boston, yang2014statistic, sienkiewicz2005statistical, huang2015cascading}, a node is a stop and any two stations served successively by a line are connected by a link. The network resulted is similar to the real system, preserving the stops distribution in the city area and making it possible to investigate general aspects such as topology, connectivity, finding hubs, clusters and elements that could cause large damage to the system if removed.

Public traffic data is extracted from the local agency {\it SPTrans} and follows the General Transit Feed Specification (GTFS) format, free to be download. This format allows traffic agencies to publish their data on a common global standard and developers to create applications that use them. Additionally, makes it easier to reproduce the steps on this research in any other city using GTFS. The format consists of up to 13 files and to recreate the network the \emph{stop\_times} is used, it contains the times that a vehicle arrives at and departs from individual stops for each trip. The sequence of stops results on an edge list to create the links and nodes for each service line. However, some links were missing in situations that we found to be connections to bus stations or terminals. Also, several stops were found for a single terminal. This characteristics resulted on multiple nodes with degree 1 and a fragmentation of the network in 11 disconnected components consisting of bus, train and subway lines. It is reasonable to assume that these stops have a connection, as trips have start/end points and most of the lines operates in cycles.

\subsection{Algorithm to group nearby stops}

\begin{figure}[h]
	\begin{center}
		\includegraphics[scale=1.1]{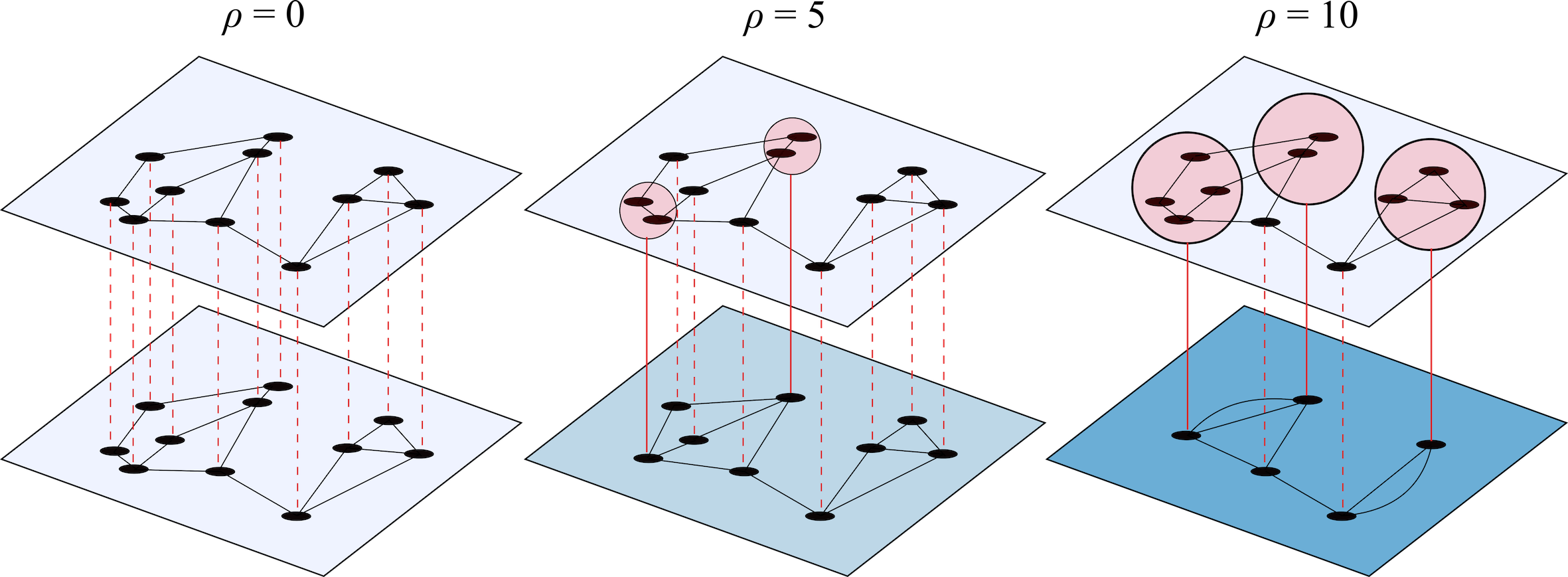}
	\end{center}
	\caption{Nodes grouped based on a distance radius $\rho$. In the illustration, for $\rho = 0$ the network is kept intact with no structural changes. At $\rho = 5$ the two pairs of nearby stops are each grouped to a single node. At $\rho =10$ each block of stops inside the radius is grouped to a single one. Nodes outside the radius are kept unchanged. Self-loops are removed and multiple links are allowed to preserve the path connectivity.
We opt to remove explicit reference to the unit of measurement, {\it meters}, on every reference to $\rho$ in order to avoid overwhelmed notation.}
	\label{model_rho}
\end{figure}

To make the network a single connected component, an algorithm was developed to group nearby stops into a single stop.
The idea is to group all stops inside a distance radius $\rho$ and represent them all as a single node in the network.
Since it is not trivial to define the ideal distance for these grouping, different values were tested within the range $0$ to $200$ meters. 
Figure~\ref{model_rho} illustrates how the system structure changes according to different distances. Every step of $\rho$ resulted on an individual network by itself, with nodes and links changing in function of stops being grouped.
From here on, wherever we refer to the distance radius $\rho$, we do not explicitly  denotes the unit of measurement, meters ($m$), just to avoid excessive notation.

The process takes place by first collecting the latitude and longitude coordinates of each stop and then all neighbours of a given stop are searched inside the radius $\rho$. 
The searching results on a dictionary with stops and the respective neighbours. So, the grouping is performed by assigning the same id to the selected stop and its neighbours. 
For instance, in Figure~\ref{model_rho} at $\rho = 5$ the two nearby stops inside the cycle receive an equal id. 
To ensure the correct formation of hubs and id assignment, a recursive search is made by also verifying the neighbours of each neighbour. 
This is required as the input file is serial and the line-up of stops could change the way groups are made.

An edge list is resulted from these steps by replacing the old id for the new one assigned during the grouping. 
At this point, the network is created and the metrics computed for each value of $\rho$, ranging from 0 to 200 meters in intervals of 5, that is, 41 different graphs are created and the metrics calculated for each one. 
These radius variation could be useful to simulate a passenger's willingness to move to nearby stops in a radius $\rho$. 
As the radius expands, the availability of lines and points increases, which can lead to a greater number of destinations and options to reach the desired location.
\begin{figure}[h]
	\begin{center}
		\includegraphics[scale=0.7]{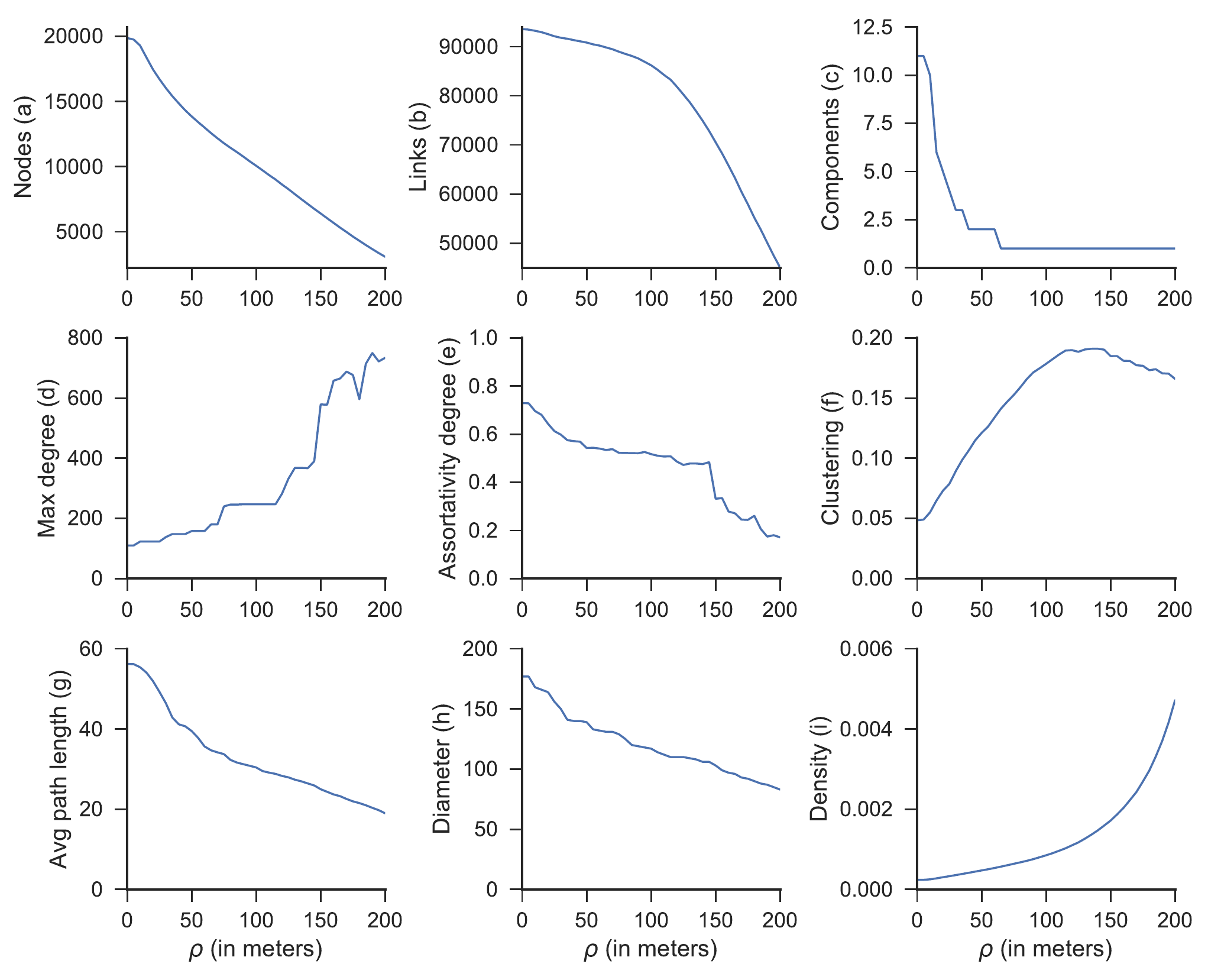}
	\end{center}
	\caption{Grid displaying statistical proprieties of the transportation system for 41 different networks created as a function of $\rho$.}
	\label{grid_stats}
\end{figure}

Indeed, by running the grouping algorithm large structural changes occurred as the distance radius increases. 
The results from this process can be seen in Figure~\ref{grid_stats}. 
The number of nodes, Figure~\ref{grid_stats}\,a, displays a strong decrease, dropping more than 80\% at the highest distance. 
A similar behavior can be seen for the links, Figure~\ref{grid_stats}\,b, but starting with a slight decrease followed by a dramatic fall. 
Even though multiple links are allowed, several self-loops were removed by nodes grouped to a single one. 
Detailed results can be seen in Table~\ref{rho_stats} where networks with significant properties changes were picked for convenience.

Starting with $11$ disconnected components, we can see in Figure~\ref{grid_stats}\,c how effective the grouping process is by the rapid drop in the number of components, even for small $\rho$ values. 
For the São PTN  Paulo, the disconnected components joined in a single one at $\rho = 65$, and remained such for higher $\rho$ values. 
The maximum degree, Figure~\ref{grid_stats}\,d, increased significantly along this process starting with a small rise in the first steps followed by a steady period in the interval $75 \leq \rho \leq 115$ and a rapid jump to its peaks at last steps. 
The line curve behaves similarly to the density, Figure~\ref{grid_stats}\,i, showing a strong correlation of 0.93 between these pair of variables.

\begin{table}[h]
	\centering
	\large
	\resizebox{\textwidth}{!}{%
		\begin{tabular}{lrrrrrrrrrr}
			$\rho$ & \emph{n} & \emph{m} & $k_{max}$ & $L_{max}$ & $\langle k \rangle $ & $\langle L \rangle$ & \emph{CC} & A & \emph{T} & \emph{D} \\ \hline
			0 & 19870 & 93607 & 110 & 177 & 9 & 56.23 & 11 & 0.73 & 0.05 & 0.0002 \\
			20 & 17461 & 92568 & 123 & 164 & 11 & 51.94 & 5 & 0.64 & 0.07 & 0.0003 \\
			40 & 14856 & 91350 & 148 & 140 & 12 & 41.16 & 2 & 0.57 & 0.11 & 0.0004 \\
			65 & 12577 & 89864 & 180 & 131 & 14 & 34.69 & 1 & 0.53 & 0.14 & 0.0006 \\
			80 & 11453 & 88543 & 246 & 125 & 15 & 32.31 & 1 & 0.52 & 0.16 & 0.0007 \\
			100 & 10070 & 86225 & 247 & 117 & 17 & 30.41 & 1 & 0.52 & 0.18 & 0.0009 \\
			120 & 8633 & 81850 & 282 & 110 & 19 & 28.30 & 1 & 0.49 & 0.19 & 0.0011 \\
			140 & 7138 & 74900 & 367 & 106 & 21 & 26.43 & 1 & 0.48 & 0.19 & 0.0015 \\
            150	& 6411 & 70539 & 579 & 103 & 22 & 24.98 & 1 & 0.33 & 0.18 & 0.0017 \\
			160 & 5690 & 65802 & 658 & 97 & 23 & 23.68 & 1 & 0.28 & 0.18 & 0.0020 \\
			180 & 4315 & 55182 & 596 & 90 & 26 & 21.53 & 1 & 0.26 & 0.17 & 0.0030 \\
			200 & 3088 & 45013 & 734 & 83 & 29 & 18.98 & 1 & 0.17 & 0.17 & 0.0047 \\
		\end{tabular}%
	}
	\caption{Network statistics for selected $\rho$ values from the 41 different graphs used in the graphs of Figure~\ref{grid_stats}. 
The the evaluated statistics are: number of nodes \emph{n}, number of links \emph{m}, maximum degree $k_{max}$, network diameter $L_{max}$, average degree $\langle k \rangle $, average path length $\langle L \rangle $, components \emph{CC}, assortativity \emph{A}, clustering (transitivity) \emph{T} and density \emph{D}.}
	\label{rho_stats}
\end{table}

\begin{figure}[h]
	\begin{center}
		\begin{minipage}[t]{0.49\textwidth}
			\includegraphics[scale=0.50]{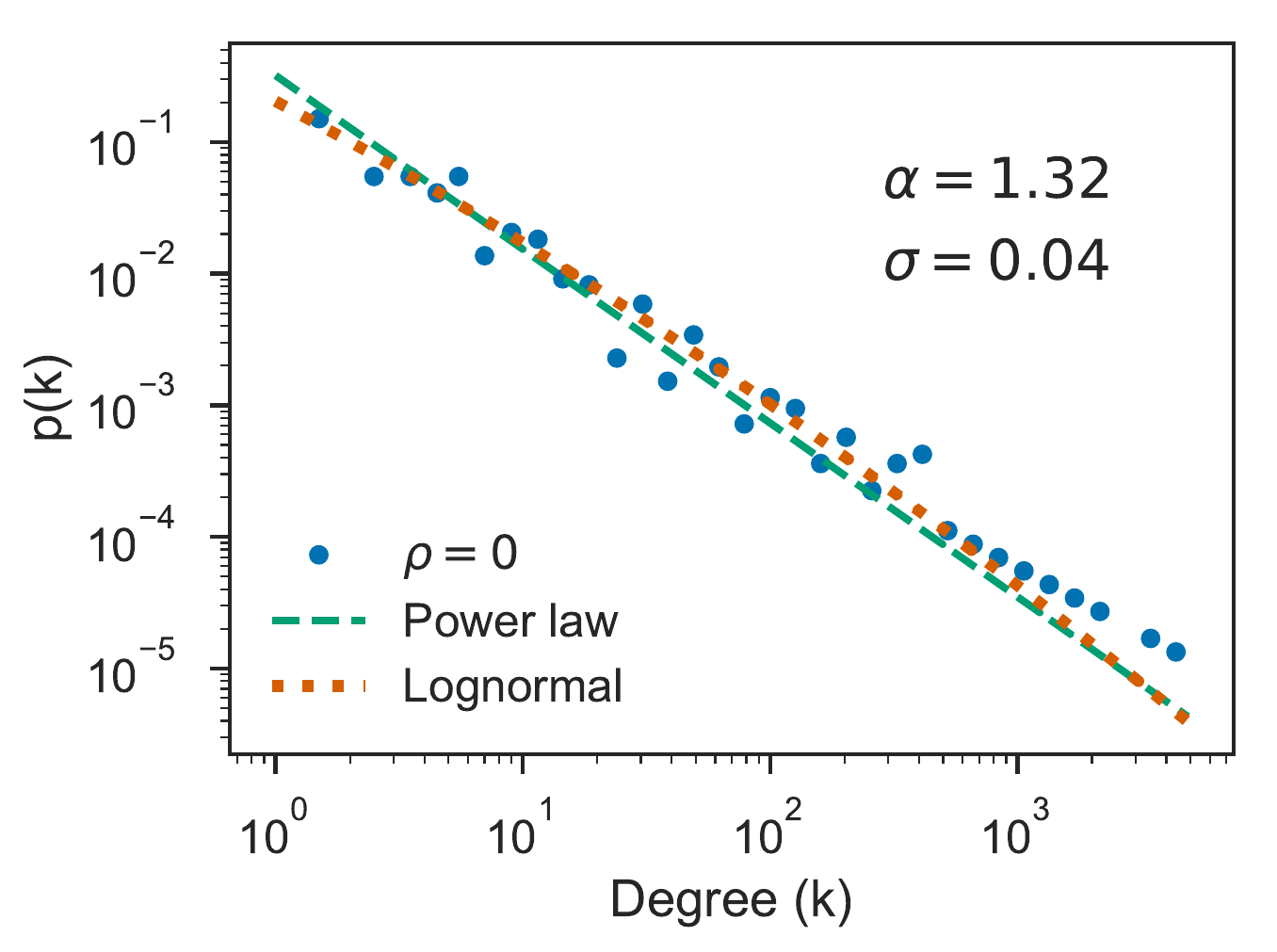}
			\includegraphics[scale=0.50]{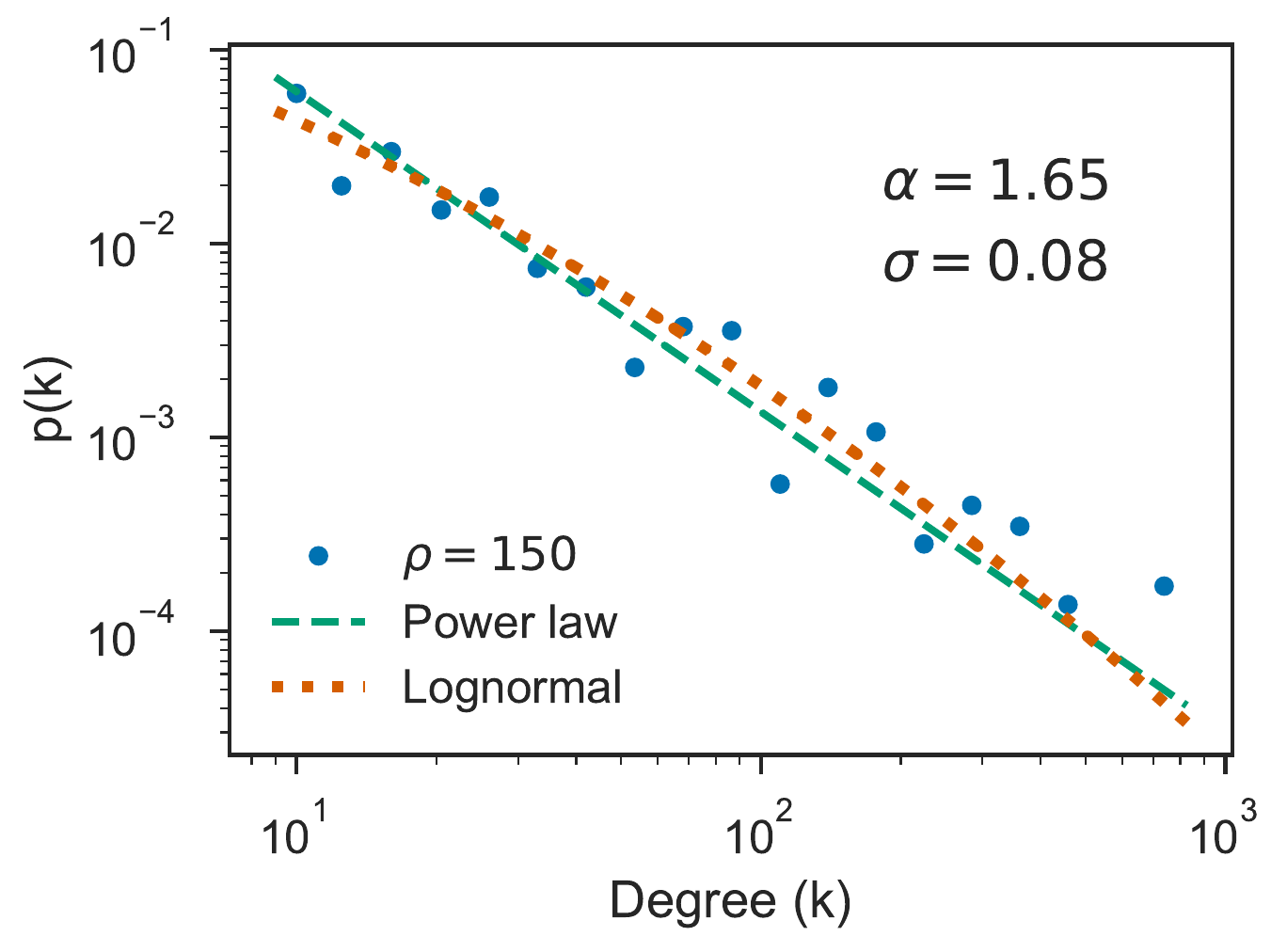}
		\end{minipage}
		\begin{minipage}[t]{0.49\textwidth}
			\includegraphics[scale=0.50]{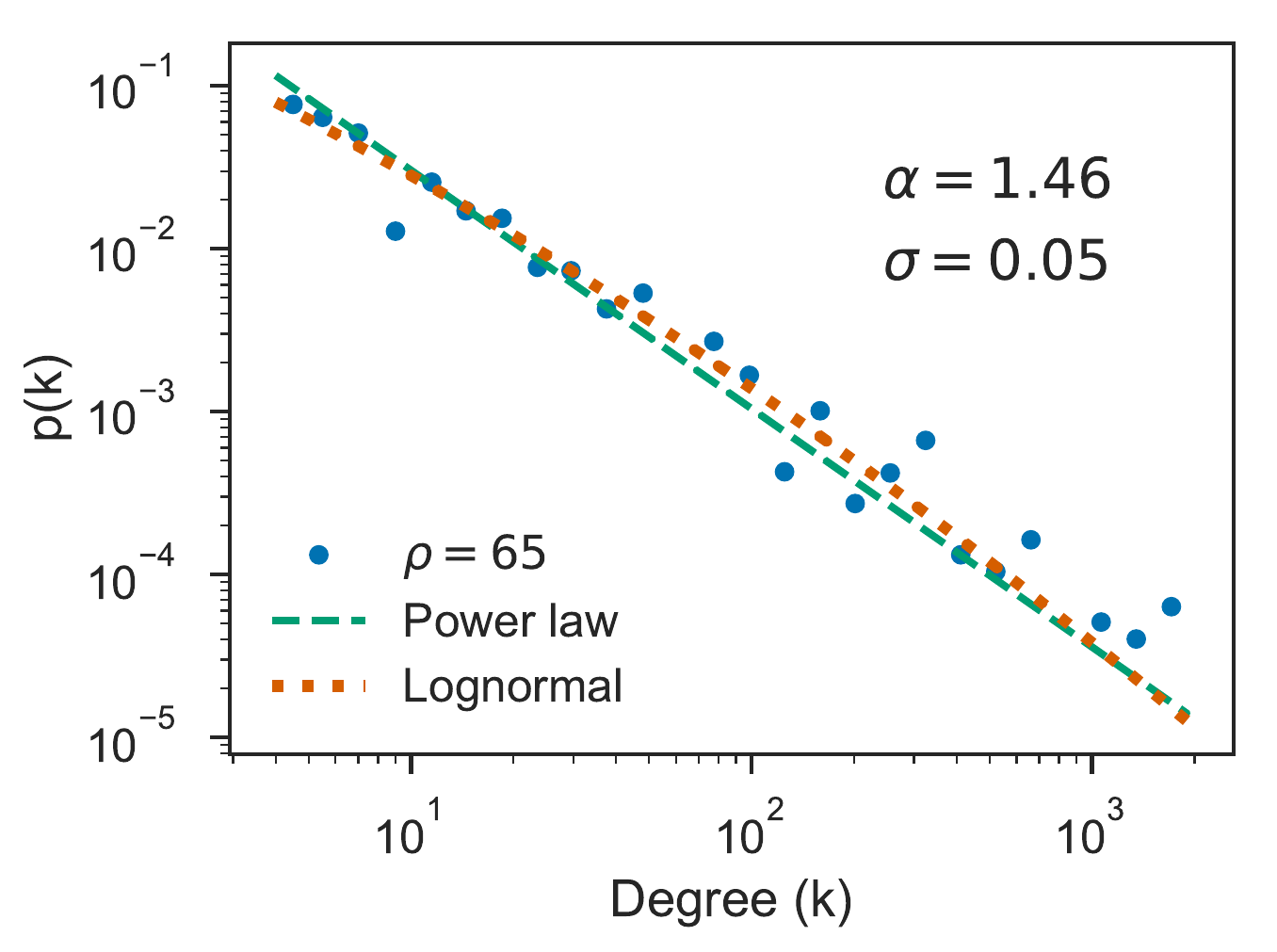}
			\includegraphics[scale=0.50]{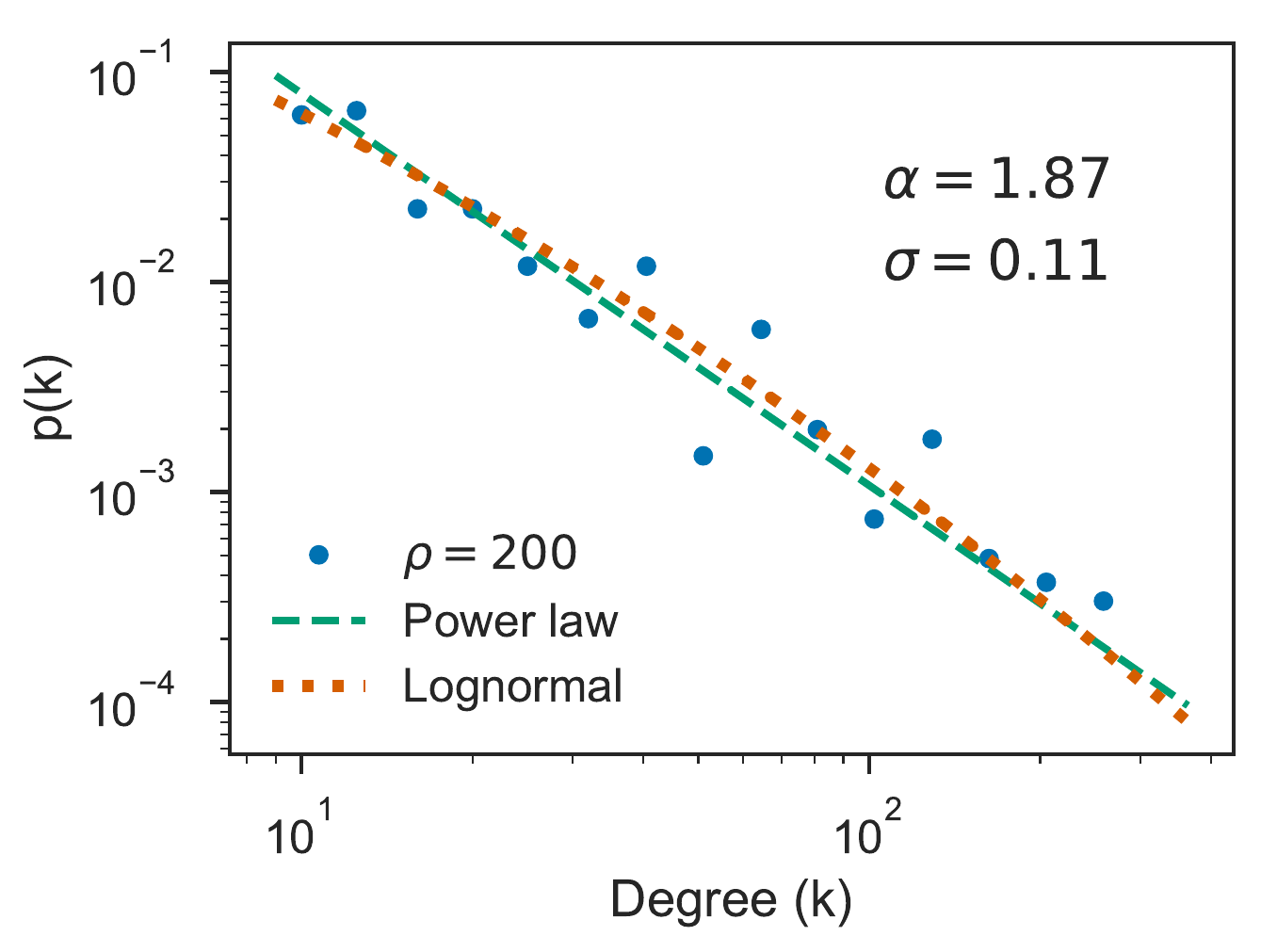}
		\end{minipage}
	\end{center}
	\caption{Degree distribution of nodes on logarithm scale with data fit in power law and log-normal. The fit was made for 4 different $ \rho $ networks. The fitted $\alpha$ parameter and its standard error $\sigma$ are shown on each plot. The data sets can be described by both distributions.}
	\label{power_law}
\end{figure}

The assortativity coefficient, see Figure~\ref{grid_stats}\,e, gradually decays from 0.73 to 0.48 at $\rho = 145$, and then has a strong fall to 0.33 just one step further, at $\rho = 150$ and keeps decreasing. 
What we see in the São Paulo PTN is a decreasing in the assortativity character with increasing $\rho$.
One expects a connection preference with nodes of the same type and high assortativity for social networks, while for technological networks, a high disassortativity is expected \cite{newman2003mixing}.
This high assortativity for the PTN is unexpected, but a possible explanation for this finding may be the large concentration of nodes with low connectivity and these nodes being linked to other poorly connected nodes. A bus stop is usually linked to only its predecessor and successor in a route. The frequency of highly connected stops is low, as can be seen in the degree distribution tail shown in Figure~\ref{power_law}. 
This decreasing value of the assortativity with $\rho$ is explained by the inclusion of before isolated or low connected stops/stations with higher connected ones, since
the assortativity was computed based on node degree, a measure strongly sensitive to the grouping process.

Nevertheless, those results are not exactly the expected ones. 
In order to check for possible computation errors in the algorithm, it was tested with random networks created with similar proportions. In 100 runs for each of random \cite{erdos1959random}, small-world \cite{watts1998collective} and scale-free \cite{barabasi1999emergence} networks, negative coefficients were obtained for assortativity, being respectively $-0.00032$, $-0.00018$ and $-0.019$. 
Being those assortativity values in the range expected, our confidence in the correctness of the values found for the São Paulo PTN was reinforced.
But for a more conclusive result, it would be interesting to compare with networks with a different heterogeneity measure. 

Reaching a peak value of 0.19 at $\rho = 140$, the clustering coefficient, Figure~\ref{grid_stats}\,f, exhibit a rapid rise followed by a slow downturn and what looks like a fall tendency. As expected, the average path length, see Figure~\ref{grid_stats}\,g, significantly decay in function of $\rho$, which may be related to the number of links dropping dramatically. Indeed, there is a positive correlation of 0.79 between the two data series. 
With a similar behavior, the system diameter displays a gradual decay reaching half of its size by the end as can be seen in Figure~\ref{grid_stats}\,h. 
Additionally, the system become more denser, Figure~\ref{grid_stats}\,i,  featuring a rapid rise after a period of slow increase, probably related to number of multiple created on grouping.

For a deeper structural analysis and connectivity comparison between subsystems, different networks for 4 values of $\rho$ were selected. The first distance $\rho = 0$ was selected as it is the initial state of the system, with no grouping performed. At $\rho = 65$ the network becomes a single component. In $\rho = 150$ there is significant drop on the assortativity with a fast increase in maximum degree. The last one, $\rho = 200$, is the highest distance tested. To observe the relationship between subsystems (bus, subway and train) within this networks, the hive plot \cite{krzywinski2011hive} was used as it provides a more comprehensive network visualization compared to traditional approach. 
The emerging pattern is resulted from the networks structure and not a layout algorithm. Nodes are mapped to and positioned on radially distributed linear axes. Edges are drawn as curved links between the axis.

\begin{figure}[h]
	\begin{center}
		\includegraphics[scale=1.3]{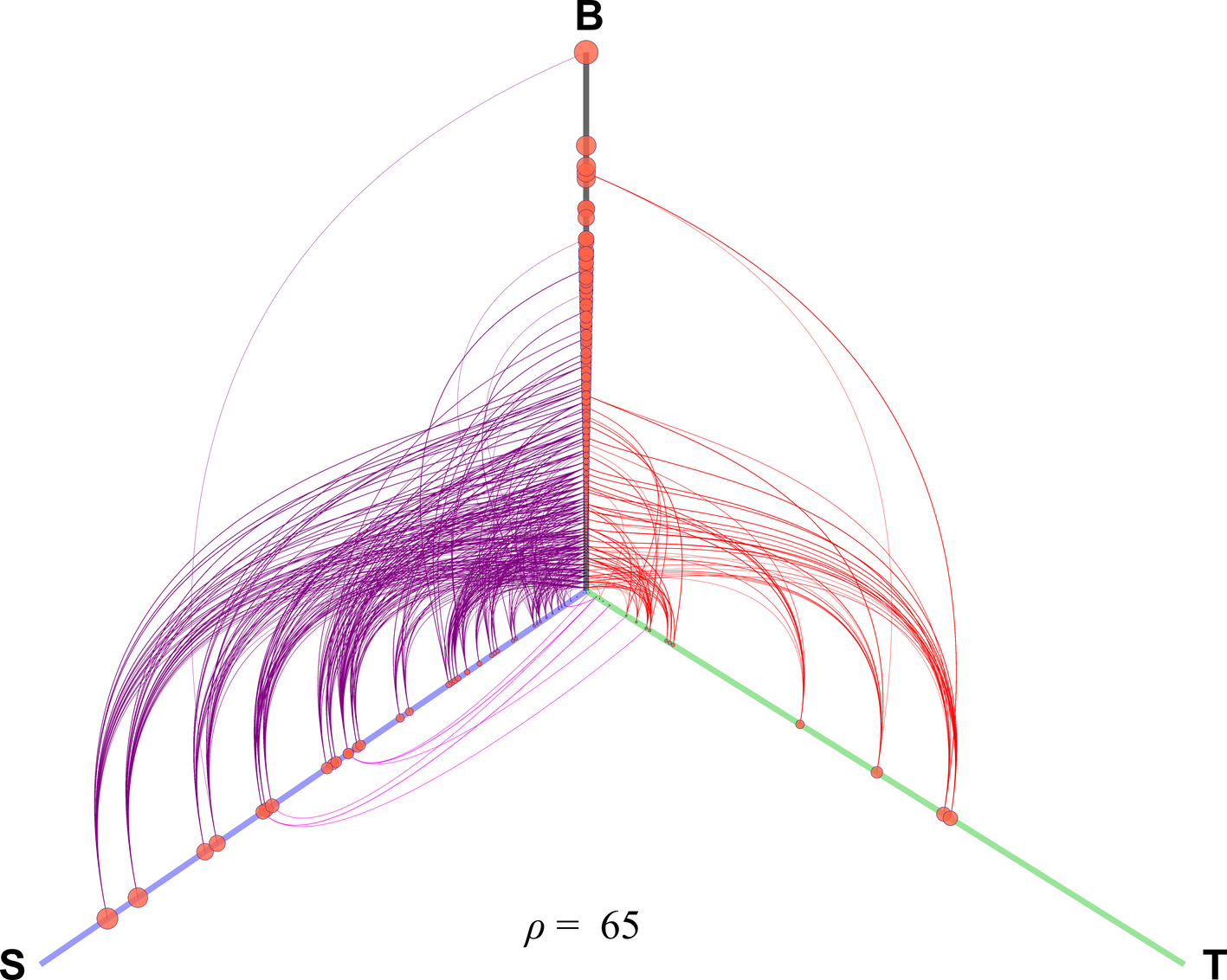}
	\end{center}
	\caption{Hive plot displaying the relationship between subsystems of the transport network at $\rho = 65$. Nodes are ordered along axis by degree and distributed according to types: B for bus stops, T for train and S for subway. The circle size represents the normalized degree. Links are drawn if source and target nodes are on different axis.}
	\label{hive_plot}
\end{figure}

Looking at the hive plot in Figure~\ref{hive_plot} it is possible to see how the connectivity between subway and bus subsystems is higher than with train subsystem. 
As nodes are normalized to the full network and not only the axis, we can see that the most connected nodes on the subway subsystem are linked to low degree bus nodes. While the bus subsystem has internally high connectivity -- internal links are not drawn but the node size represent its degree -- the subway nodes shows high external connectivity. A similar behavior was found between train and bus subsystem but on a lower scale. Between subway and train subsystem only few links are presented as observed on the real system with a small number of transfer stations connecting them.

\begin{figure}[h]
	\begin{center}
		\begin{minipage}[t]{0.32\textwidth}
			\includegraphics[scale=0.76]{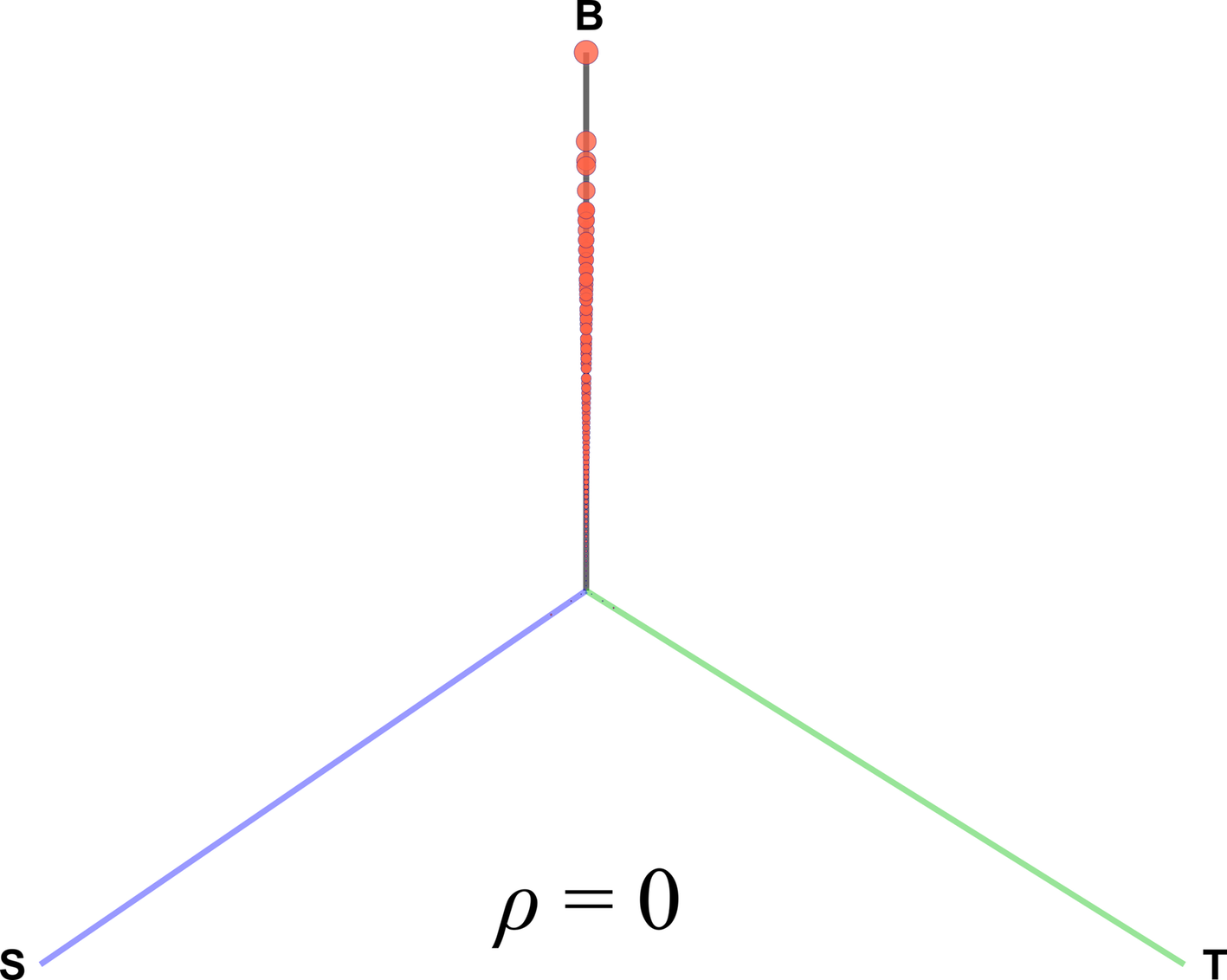}
			\centering \small
		\end{minipage}
		\begin{minipage}[t]{0.32\textwidth}
			\includegraphics[scale=0.76]{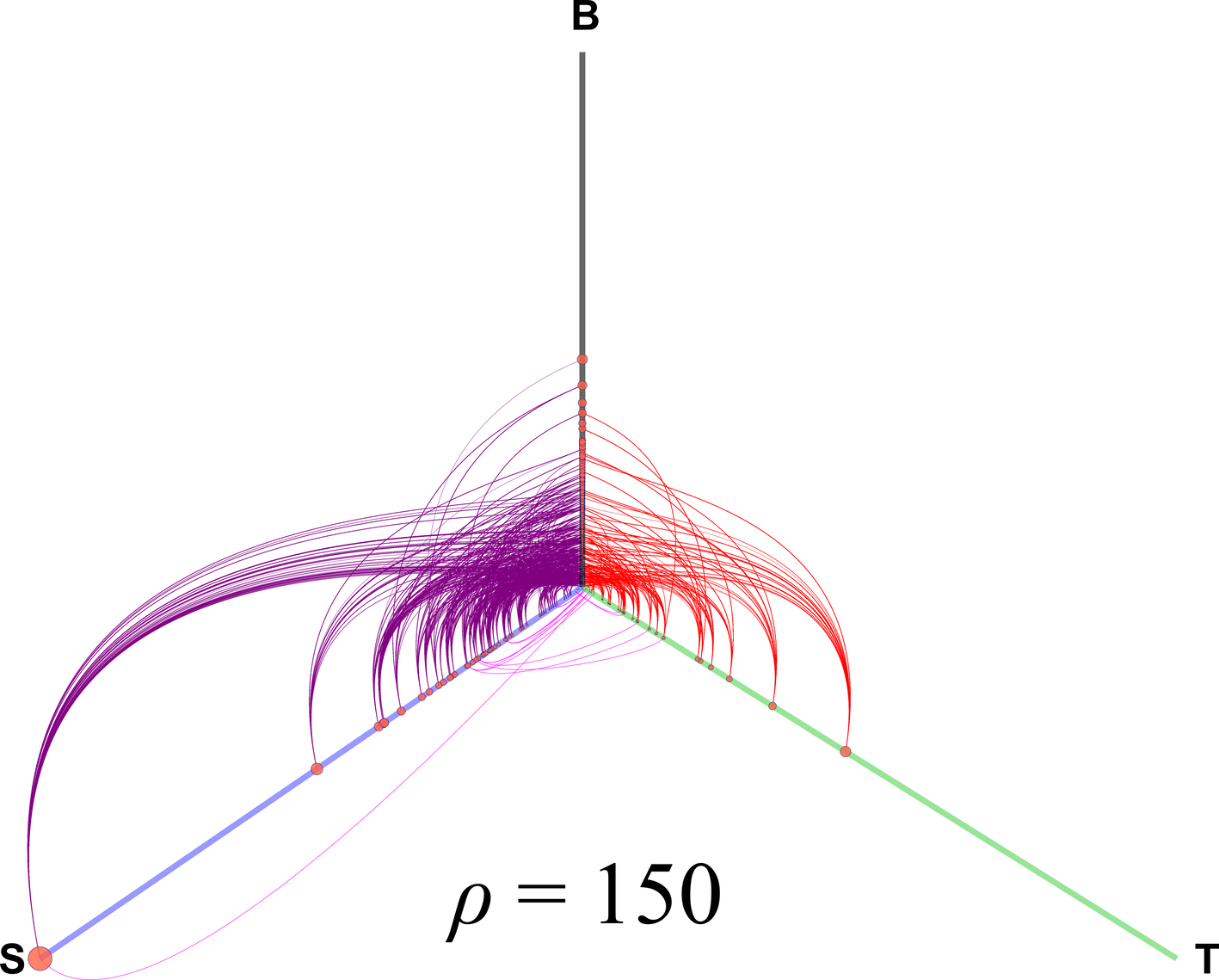}
			\centering \small
		\end{minipage}
		\begin{minipage}[t]{0.32\textwidth}
			\includegraphics[scale=0.76]{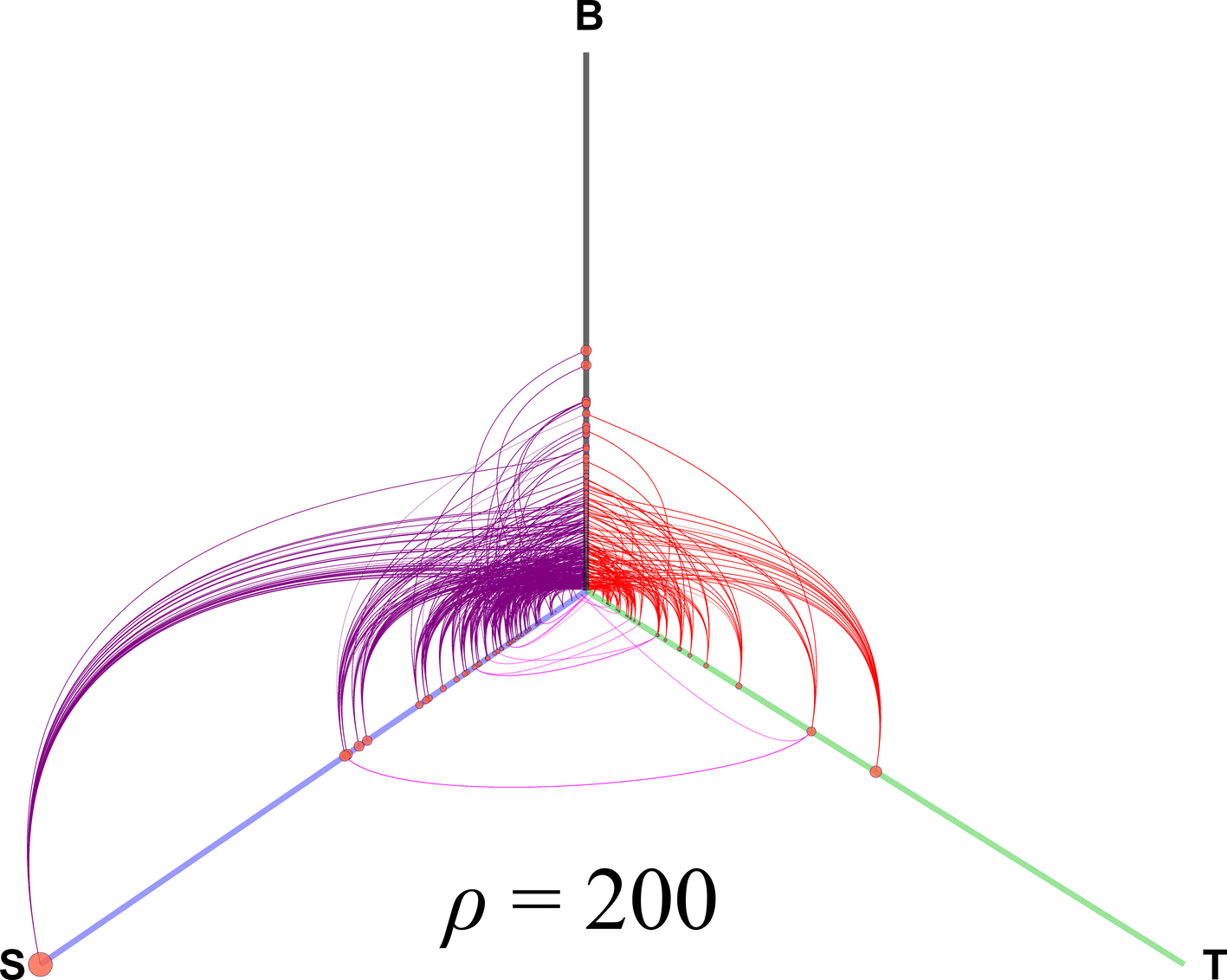}
			\centering \small
		\end{minipage}
	\end{center}
	\caption{Hive plot displaying the relationship between subsystems of the transport network for $\rho = 0$, 150 and 200. At $\rho = 0$ there are no links as the system starts from a state with 11 disconnected components and missing links between subsystems. $\rho = 150$ and 200 displays a highly concentrated state.}
	\label{hive_grid}
\end{figure}

In Table~\ref{assortativities} at $\rho = 65$ we can see that this subsystems have a positive correlation as found by the assortativity degree computed between them. 
Results for other values of $\rho$ can also be observed in the table. 
In Figure~\ref{hive_grid}, networks for other values of $\rho$ were also plotted in the same visualization style. Drawing the network at $\rho = 0$ did not result in any pattern due to missing links connecting the subsystem as stated before. We can see clear structural changes over time on the different networks, the system transit from a null external connectivity state at $\rho = 0$ to a well distributed one at $\rho = 65$, however, it reached highly concentrated states at $\rho = 150$  and 200. If we evaluate this properties keeping in mind the system robustness, $\rho = 65$ seems to be a good candidate distance to group stops as hubs are better distributed along axis and between subsystems.

In Table~\ref{assortativities} we can see the detailed assortativity results between the network subsystems. At $\rho = 0$ only a global value could be computed, at this state there is no connectivity between subsystems due to the 11 disconnected components and missing links. At $\rho = 150$ a negative value was found for train and subway, however, on the next plot a strong positive value was obtained. A clue of what could have caused this behavior is the maximum degree, it jumped fast at $\rho = 150$ reaching maximum value near $\rho = 200$.

\begin{table}[h]
	\centering
	\small
	\begin{tabular}{lrrrr}
		\hline
		$\rho$ 	    & Global	& B - T     & B - S  & T - S \\ \hline
		$\rho=0$    & 0.72	&  -        &  -   	&  -  	\\
		$\rho=65$   & 0.53   & 0.40      & 0.29 	& 0.41	\\
		$\rho=150$  & 0.33	& 0.43	    & 0.02 	& -0.13	\\
		$\rho=200$  & 0.17	& 0.35	    & 0.39 	& 0.89	\\
	\end{tabular}
	\caption{Table displaying assortativity degree between sub networks for different values of $\rho$. At $\rho=0$ there is no connectivity between the subsystems.}
	\label{assortativities}
\end{table}

A typical analysis in complex networks is to verify if the degree distribution of nodes follows a power law, which is described by the form $P(k) \propto k^{\alpha}$. 
The constant $\alpha$ is known as the scaling exponent, which typically appears in the range $ 2 < \alpha < 3 $. 
In practice, few phenomena obey power law for all values of \emph{k}, being more applied to values greater than some minimum \emph{k}, in these cases is common to say that only the tail of the distribution follows a power law \cite{clauset2009power}.

Adjust a power law to a set of data may be tricky. So, we decide to use the
the \emph{Powerlaw} Python package \cite{alstott2014powerlaw} was used to plot and fit the data. It is an open source and free library that allows the use of different types of distribution to test the data sets. The input is the histogram of the $k$ distribution and values with zero are deleted during fit as axes follow a logarithmic scale. The analysis is done for the 4 networks mentioned before, see Figure~\ref{power_law}. Fitting between power law and log-normal are almost indistinguishable, that is, both distributions can be used to describe the data sets behavior.

We can see a tendency of $\alpha$ increasing in function of $\rho$. A statistical test was made to evaluate if the distribution could be better described by a power law or log-normal. For the four networks ($\rho = 0$, $65$, $150$ and $200$) we calculated the ratio of the probability ($R$) between the two distributions obtaining respectively $-3.52$, $-2.85$, $-2.73$ and $-1.2$. A positive $R$ indicates that the data is more likely to be described by a power law and negative for a log-normal. The significance of $R$ is respectively $0.12$,  $0.14$,  $0.14$ and  $0.32$. The results showed that all the data sets could be better described by a log-normal function \cite{}.

To evaluate if the networks fall in the small-world class we used a quantitative categorical definition \cite{humphries2008network}. Such test implies comparing the average path length and the clustering coefficient with a random network of the Erdös and Rényi (ER) \cite{erdos1959random} type with the same number of nodes \emph{n} and links \emph{m}. 
The empirical network constructed from the São Paulo PTN with $\rho$ is expected to displays an average path length $L_{\rho}$ similar to a random graph $L_{rand}$ and a clustering coefficient $C_{\rho}$ greater than an equivalent random graph $C_{rand}$. By taking the quotients $\gamma_{\rho} = C_{\rho} / C_{rand}$ and $\lambda_{\rho} = L_{\rho} / L_{rand}$ we can find the small-world-ness of the networks by $S = \gamma_{\rho} / \lambda_{\rho}$. This implies $\lambda_{\rho} \geq 1$ and $\gamma_{\rho} \gg 1$, so a network is said to be small-world if $S > 1$. Detailed results from this test can be seen in Table~\ref{small_worlds}.

\begin{table}[h]
	\centering
    \large
    \resizebox{\textwidth}{!}{%
    	\begin{tabular}{lrrrrrrrrr}
    		\hline
    		$\rho$ & \emph{n} &\emph{m} & $C_{\rho}$	& $C_{rand}$ & $L_{\rho}$	& $L_{rand}$ & $\gamma_{\rho}$ & $\lambda_{\rho}$ & \emph{S} \\ \hline
            
            $\rho=0$ &  19870 &  93607 &  0.048 &  0.0005 &  56.23 &    6.55 &  103.14 &    8.58 &  12.02 \\
            $\rho=65$ &  12577 &  89864 &  0.141 &  0.0011 &  34.69 &    5.03 &  127.38 &    6.90 &  18.47 \\
            $\rho=150$ &   6411 &  70539 &  0.185 &  0.0035 &  24.98 &    3.91 &   53.60 &    6.39 &   8.39 \\
            $\rho=200$ &   3088 &  45013 &  0.166 &  0.0091 &  18.98 &    3.29 &   18.27 &    5.76 &   3.17 \\
    	\end{tabular}%
    }
	\caption{Table showing small-world test for the 4 networks compared to a random network of Erdös and Rényi type. An $S > 1$ means that the network falls in the small-world class.}
	\label{small_worlds}
\end{table}

All the networks displayed a positive quotient \emph{S} greater then 1 meaning that the small-world property may probably be present on the system in study here. $C_{\rho}$ and $L_{\rho}$ are considerably larger then all random networks. The $L_{\rho}$ is not similar to the real network, but it is very difficult reproduce the \emph{L} criterion as the ER model is constructed by assigning each unique link to a pair of nodes with uniform probability, disregarding how sparse the network is. This makes the graph very close to a connected component due to links and nodes ratio, as can be seen with the low $L_{rand}$ for all networks.

\section{Simulated attacks to nodes and links}\label{attacks}

Robustness is an important property in infrastructure PTN due to the crucial role it plays in cities. 
A PTN involves the transport of people and a critical aspect of its operation is to continue connecting people to their destinations in face of perturbations. 
In order to test the robustness of São Paulo PTN we perform \emph{attacks}, meaning  the simulation of possible system failures, by removing nodes or links.
In practice, such attacks could be mechanical failure caused by power outage, broken vehicles, emergency maintenance; accidents on roads involving system's vehicles or passengers; road obstruction by planned or unplanned construction, temporarily closure; natural disasters and so on.
Different attacks scenarios are used to evaluate the robustness, either selecting stops or links, randomly or directed, being the number of components evaluated along the process. 
These strategies are described as follows:

\begin{enumerate}
    \item \textbf{Random nodes} consists in the removal of nodes with uniform probability.
    \item \textbf{Probabilistic targeted nodes} consists in the removal of nodes with probability proportional to their degrees, in roulette schemata. 
    \item \textbf{Deterministic targeted nodes} consists in the removal of nodes from the highest degree to the smaller ones. If there are more than one node with the same degree, both are removed in sequel. 
    \item \textbf{Random links:} consists in the removal of links with uniform probability, independently of their multiplicities.
    \item \textbf{Probabilistic targeted links} consists in the removal of links with probability proportional to their multiplicities, in roulette schemata. 
    \item \textbf{Deterministic targeted links} consists in the removal of links from the highest multiplicity to the smaller ones. 
Multiplicity in this context reflects how many times that path is used in the network.
If there are more than one link with the same multiplicity, both are removed in sequel. 
\end{enumerate}

\begin{figure}[th]
	\begin{center}
		\includegraphics[scale=0.52]{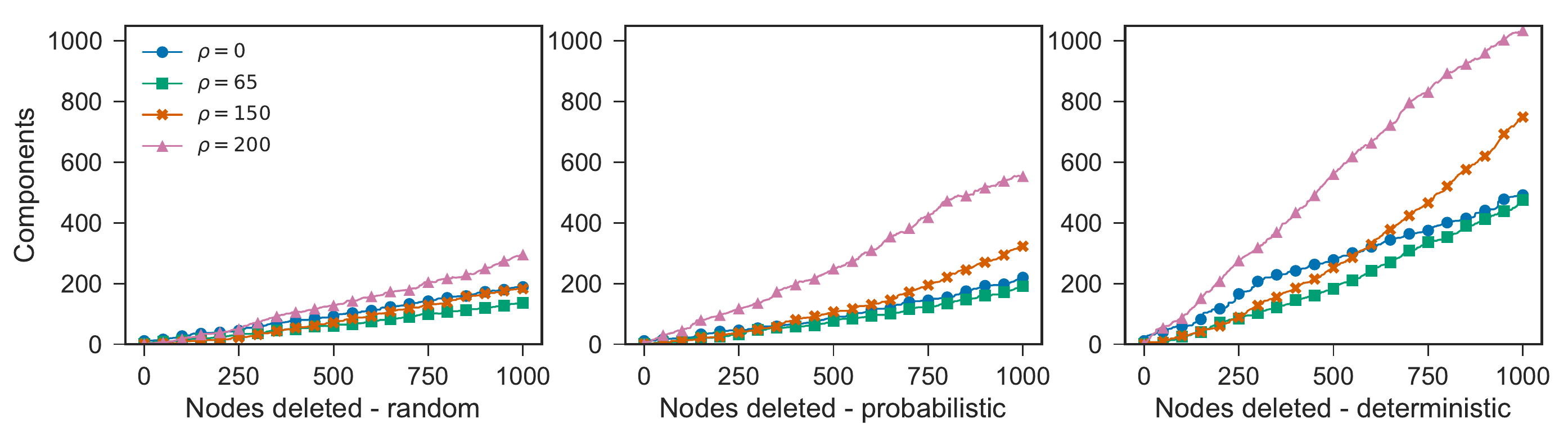}
	\end{center}
	\caption{Impact on the number of components in function of nodes removed on the random,  probabilistic and deterministic strategies for the 4 selected networks, $\rho = 0$, 65, 50 and 200.}
	\label{attack_nodes_comp}
\end{figure}

We performed these 6 different strategies attacks to the same 4 preselected networks indicated in table~\ref{assortativities}, namely, the networks with $\rho = 0$, $65$, $150$ and $200$.
The robustness analysis presented here shows how the networks number of components and average path lengths respond in the different scenarios and capture which one can cause a higher damage to the system. 
Although a single component is obtained at $\rho = 65$, it is important to test the system in previous states (with disconnected components) to investigate whether the grouping influences or not in the system robustness.

Figure~\ref{attack_nodes_comp} shows the number of components as function of the number of nodes deleted for strategies 1 to 3, applied to the São Paulo PTN with $\rho = 0$, 65, 150 and 200.
It is noticeable that the impact of removing the most connected nodes using the deterministic strategy (Figure~\ref{attack_nodes_comp}\,c) is much larger than randomly removing the nodes (Figure~\ref{attack_nodes_comp}\,a). 
At $\rho = 0$ the network already starts in a fragmented state while the networks with $\rho \geq 65$ start from a single component. 
The curves for the number of components have a similar behavior for all attack strategies, with $\rho = 200$ showing a fastest rise. 
This maybe due to the rich connected nodes removed in networks with $\rho = 200$, when there are nodes with degree greater than 700, as shown in Figure~\ref{grid_stats}\,d.
Additionally, for $\rho=200$ there is a smaller number of nodes, and $\approx 30\%$ nodes were removed from the total of 3088, much larger then the other networks tested. 

\begin{figure}[th]
	\begin{center}
		\includegraphics[scale=0.52]{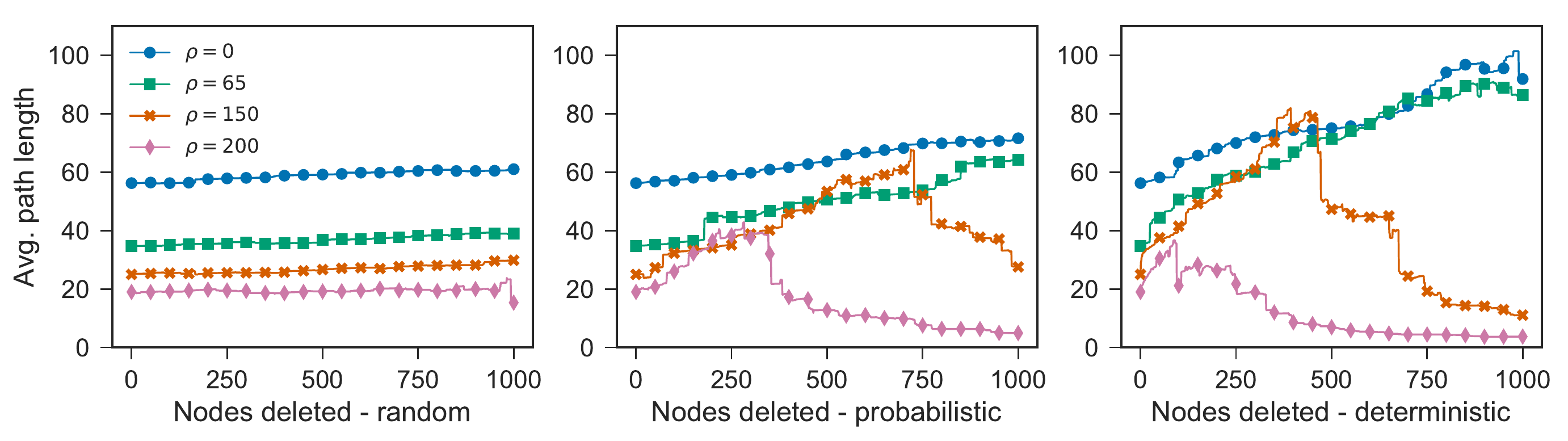}
	\end{center}
	\caption{Impact on the average path length in function of nodes removed on the random,  probabilistic and deterministic strategies for the 4 selected networks, $\rho = 0$, 65, 50 and 200.}
	\label{attack_nodes_path}
\end{figure}

Another important system's property is the average path length between two random points. 
In  Figure~\ref{attack_nodes_path} we notice that it slowly increases as nodes are removed in the random strategy. 
A similar pattern was found for $\rho = 0$ and 65 in the probabilistic strategy, but for $\rho = 150$ and 200 the measure displays a faster increase before a strong downturn in different points, following by a decrease tendency. 
The deterministic mode presents similar curves to the probabilistic mode, but with faster and more dramatic drops. 
Farther, the networks of $\rho=0$ and 65 start a little drop by the end of simulation.

The dramatic fall in $ \rho = 150 $ close to the 500th node removed on deterministic strategy displays an interesting result: the path length keeps rising until reaching a critical point where removing more nodes makes the network to collapse and strongly drop paths. 
Even though in denser networks this happens faster as seen with $\rho = 200$, for $\rho = 150$ we have an intermediate configuration where the network seems to be capable of observe a high number of deleted nodes before reaching a turn point. 
Perhaps this result is related to the number of components being fragmented much faster when compared to the other networks, see Figure~\ref{attack_nodes_comp}\,b  and c. 
Another possible cause for the behavior of these falls is the presence of path redundancies as the grouping algorithm transformed several nearby stops/stations into one. 
Those existing old paths became multiple links, which do not contribute to the path length.

\begin{figure}[th]
    \begin{center}
        \includegraphics[scale=0.52]{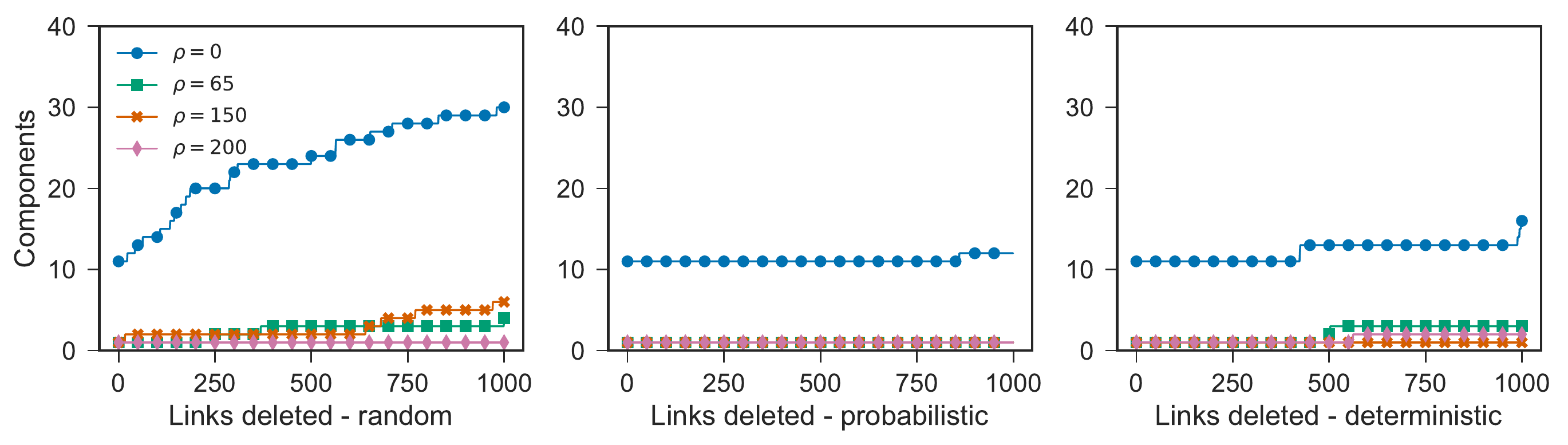}
    \end{center}
    \caption{Impact on the number of components in function of links removed in the random,  probabilistic and deterministic strategies for the 4 selected networks, $\rho = 0$, 65, 50 and 200.}
    \label{attack_links_comp}
\end{figure}

Next, we analyze the number of components and the average path length between two random points as function of the number of links deleted for strategies 4 to 6, applied to the São Paulo PTN with $\rho = 0$, 65, 150 and 200.
It is expected that attacks to the links wold cause a lower impact on the system due to the higher number of links compared to nodes. 
It is also important to clarify that the links weights are assigned based on the path multiplicity in the network, that is, the number of times that path between two points is used by the lines. 
This option to assign a weight to the links based in their multiplicity was done because actual flows of people and vehicles are not easily available. 
However, this weight is enough to evaluate the link importance in the robustness context analysis, as removing it can influence directly in the number of lines covering the respective path.

Figure~\ref{attack_links_comp} shows the number of components for the removal of links strategies.  
Two observations are due here. 
First, the $\rho = 0$ network has an unexpected higher fragmentation if compared with the $\rho =$ 65, 150 and 200 since for the removal of nodes strategies (Figure~\ref{attack_nodes_comp}) the order is reversed, as $\rho =$ 200 has the higher fragmentation if compared whit $\rho =$ 150, 65 and 0. 
These behavior could be a result of fewer multiple links and to the fact of most nodes being poorly connected in $\rho = 0$ network, as can be seen by the low average degree in Table~\ref{rho_stats}. The other networks remained stable most the time with a slight increase at the end of simulation.
Second, the random removal of links strategy (Figure~\ref{attack_links_comp}\,a) has higher fragmentation than the probabilistic (Figure~\ref{attack_links_comp}\,b) and deterministic (Figure~\ref{attack_links_comp}\,c) removal strategies, in opposite behavior of nodes removal strategies (Figure~\ref{attack_nodes_comp}). 

\begin{figure}[th]
	\begin{center}
		\includegraphics[scale=0.52]{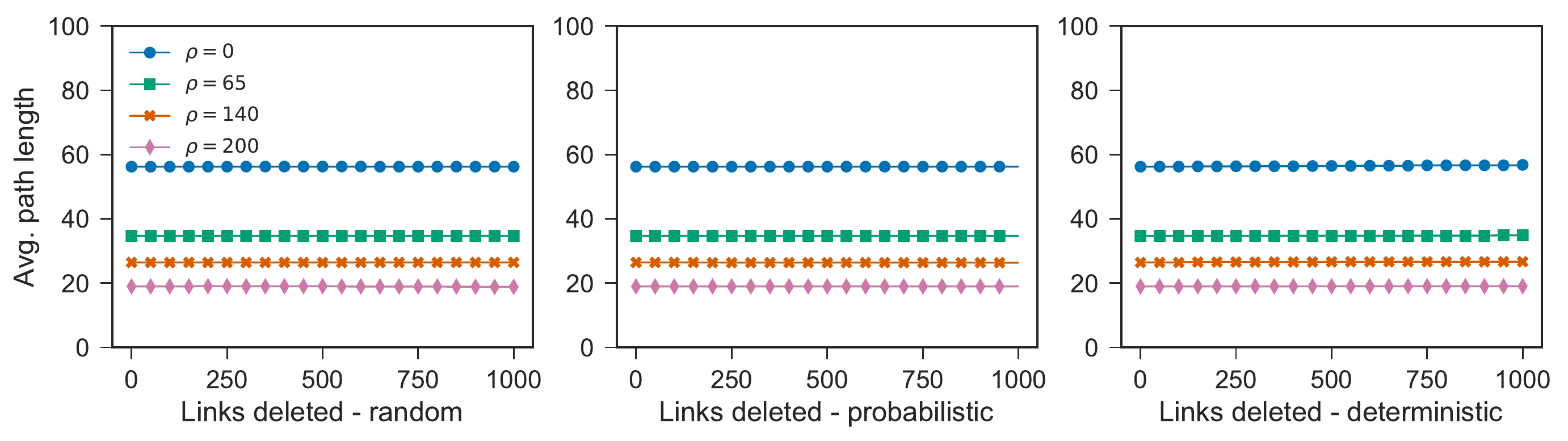}
	\end{center}
	\caption{Impact on the average path length in function of links removed on the random,  probabilistic and deterministic strategies for the 4 selected networks, $\rho = 0$, 65, 50 and 200.}
	\label{attack_links_path}
\end{figure}

Looking at the average path length in Figure~\ref{attack_links_path} we can see that it remained stable for all the strategies and networks. The average distance does not increase with the absence of links removed, a result probably related to their multiplicity. It is likely that this behavior is related to the small-world propriety which presents a high clustering index for a relatively short average path length. However, it would be interesting to use a different centrality measures to select links for deletion as the multiplicity did not prove to be relevant as an impact factor for connectivity.

Overall, the removal of links strategies had small impact for all of other measures such as density, diameter, clustering, assortativity and degree. 
It seems that links are more robust than nodes, specially on probabilistic and deterministic strategies, as illustrated for the number of components in Figure~\ref{attack_links_comp}. Making a parallel with the real system, we can look at this results as the interruption of a single line compared to large bus terminals and nearby subway stations. 
By removing hubs, dozens of lines will be interrupted, consequently the nearby stops serving as transfers can also be affected.

\section{Discussion}

The grouping algorithm showed interesting results that could be used to simulate user's willingness to walk for transportation access, but the current study was not specifically designed to evaluate this property. The algorithm results proved to be an effective method to fill the missing links from GTFS data, a goal obtained successfully at a radius of 65 meters. Indeed, this finding is a hint for similar works dealing with the same data structure in the future. For a more realistic analysis of user's willingness, it would be interesting to incorporate different link weights such as travel time, financial cost, number of transfers and so on. 

In the meantime, strong changes in the system's structure could be observed by testing different $\rho$ distances, under removed nodes and links attacks.
The fast drop of links close to $\rho = 100$ (Figure~\ref{grid_stats}\,b) is a result of several self-loops removed which increases in function of nodes grouped. 
Further work needs to be done to establish whether the assortativity by degree holds for the network. 
A hint for this result could be in Figure~\ref{hive_grid} at $\rho = 0$ where it can seen the majority of nodes of type bus. 
They are connected internally and most concentrated on the low degree range. 
A bus stop is usually linked to only two more stops on a route, meaning the network is very sparse due to spatial constrains. 
The frequency of highly connected stops is low and they are usually linked to similar rich connected nodes.

S\~ao Paulo PTN is characterized to the existence of many bus stop and fews terminals. We think this feature explain the high assortativity found in this public transportation network. In peripheral areas this pattern is even more recurrent.   The more polycentric the city become the more terminals arise to connect different centers.  Terminals are important to promote alternative shortcuts to those travel between different peripheral district. However, public transportation in S\~ao Paulo is more directed to the central regions of the city. More jobs are offered near downtown. Hence, in S\~ao Paulo land are more expensive in those areas which offer a better transport options and jobs. In the case of S\~ao Paulo the transport network system penalize poor people who lives in peripheral district and spend more time traveling which is also costly. So, such PTN help to promote income inequality.

The degree distribution of nodes on logarithm scale (Figure~\ref{power_law}) nicely fits in a  power law, as well as in a log-normal distribution, meaning that both distributions able to describe the data, although a slightly better fit was found for a log-normal. 
However, the discussion if a data set follows or not a power law is not simple. 
Even the use of some test to find the best fit is often insufficient for a conclusion. Systems found in the real world have noise, so that a power law with the perfection of a theoretical distribution rarely occur. 

By simulating node attacks to the system we could see a fast fragmentation when the most important nodes are targeted (Figure~\,\ref{attack_nodes_comp}).
If we take this behavior as an indication for vulnerability, the networks of $ \rho = 0$ and 65 appear to be slightly better then 150 and strongly more robust then $\rho=200$. 

The fast drop in average path length for $\rho = 150$ and 200 (Figure~\ref{attack_nodes_path}) can be related to the high proportion of nodes deleted, at this state removing 1000 nodes represents $\approx 25\%$ and $30\%$ respectively, which may cause dramatic impact in the network structure. 
We can also see a strong fall in the number of links for $\rho$ greater than 100 (Figure~\ref{grid_stats}\,b). 
Some nodes in the network play the role of anchor, allowing bridges between parts of the network and when they are removed, the path length increases since a new path results in more steps. 

Even though the network becomes significantly denser after $\rho = 150$, the higher proportion of links does not mean greater robustness as most of the links could be multiple for the same path. 
Additionally, distant nodes at periphery areas for São Paulo PTN are usually far from each other and increasing the radius does not result in any change to the collapsed network.
An opposite outcome occurs in the city center, where a better transport infrastructure is available.

A deeper analysis on removal of links strategy would be required for a better comprehension of the simulations outcomes.
It looks like links are more robust but this could be an outcome directly influenced by the algorithm used as removing a link on a path with multiple links between to given nodes would not change the average path length. 
The multiplicity weight used to probabilistic sample the links in the removal of links attacks is a simple frequency based calculation.
A variation of this method would be remove the role path instead of only one link -- what is the same of recreating the network without multiple links -- would probably fragment the network much faster and would give an idea of a long range area damage. 
It is also important to point that the two type of targets are distributed in different proportions in the networks, on a scale of $\approx 5 $ links for each node, a proportion that increases with $\rho$. 
For a more balanced comparison, it would be interesting to remove links and nodes in terms of proportion to the total, such as $20\%$ of links and nodes, depending on the kind of attack considered.

\section{Conclusion}

The grouping algorithm proved to be an effective method to fill missing links, goal obtained with the single component at radius of $\rho = 65$ meters. 
Additionally, this process showed an useful outcome to evaluate mobility public policies, where distance can be used to simulate the user's willingness to move around and access different transport options in relation to their original position.

São Paulo's PTN has a relatively short average path length in relation of its diameter, most stops are poorly connected and degree distribution that can be described by both power law and log-normal. 
We could not find evidences of scale-free properties on the tested networks but all of them fall in the small-world class according to the test performed. 
The hive plots displayed strong assortativity degree between the PTN subsystems, showing that maybe there is a preference mechanism based on connectivity governing how nodes are linked.

The deterministic strategy proved to be more impacting in most of the tested scenarios, followed by probabilistic and at least random. 
If fragmentation is considered as a criterion of robustness, the networks with $\rho = 0 $ showed to be more robust for removal of nodes attacks, while networks with $\rho = 200$ showed to be more robust if the path length is considered.
However, it was clear that removing links cause significantly less damage than removing  nodes, an outcome probably related to the multiplicity of links resulted from the grouping process. 

\section*{Acknowledgement}
SFS thanks National Council for the Improvement of Higher Education (CAPES) for the financial support during the realization of this work.

\bibliographystyle{ieeetr}
\bibliography{ptn}

\end{document}